\documentclass[aps,prl,twocolumn]{revtex4-1}
\pdfoutput=1
\usepackage{hyperref}
\usepackage{graphicx}
\usepackage{color}

\begin{document}

\title{Geometrically induced modification of surface plasmons\\ in the optical and telecom regimes}

\author{M. L. Nesterov,$^{1,2,*}$ D. Martin-Cano,$^1$  A. I. Fernandez-Dominguez,$^{1,\dag}$\\
E. Moreno,$^1$  L. Martin-Moreno,$^3$ and F. J. Garcia-Vidal$^{1,\ddag}$}

\address{
$^1$Departamento de F\'{i}sica Te\'{o}rica de la Materia Condensada,
Universidad Aut\'{o}noma de Madrid, E-28049 Madrid, Spain\\
$^2$A. Ya. Usikov Institute for Radiophysics and Electronics NAS of Ukraine,\\
12 Academician Proskura Street, 61085 Kharkov, Ukraine\\
$^3$Instituto de Ciencia de Materiales de Arag\'on (ICMA)
and Departamento de F\'{i}sica de la Materia Condensada,\\
CSIC-Universidad de Zaragoza, E-50009 Zaragoza, Spain\\
$^*$Corresponding author: nesterovml@gmail.com\\
$^\ddag$Electronic address: fj.garcia@uam.es
}

\date{\today}

\begin{abstract}
We demonstrate that the introduction of a subwavelength periodic
modulation into a metallic structure strongly modifies the guiding
characteristics of the surface plasmon modes supported by the
system. Moreover, it is also shown how a new type of a tightly confined surface plasmon polariton mode can be created by just milling a
periodic corrugation into a metallic ridge placed on top of a metal surface.
\end{abstract}

\maketitle

Photonics based on the exciting capabilities of surface plasmon polaritons (SPPs) has become a very active area of research during the last decade~\cite{Barnes2003,Zayats2005,Maier2007}. Strong localization of electromagnetic (EM) fields and the building up of ultra-small SPP-based waveguides~\cite{Buckley2007,Bozhevolnyi_Book} have been achieved thanks to the subwavelength (the term ``subwavelength'' refers to the vacuum wavelength) nature of the SPP fields, thus enabling a great variety of applications in optics~\cite{Ditlbacher2002,Ebbesen2008}.

However, in order to fulfill the potentialities of light guiding
based on SPP excitation, it is convenient to search for effective
ways to control and tune the propagation characteristics of SPP
modes. The concept of geometrically-induced SPPs~\cite{Pendry2004,Hibbins2005,deAbajo2005,Maier2006,Williams2008,Fernandez-Dominguez2009}
 (also named {\it spoof} SPPs) has proven
to be very powerful in tailoring the dispersion relation of the
propagating surface EM modes in the microwave and terahertz
regimes. The aim of this Letter is to transfer the spoof SPP
concept to the optical and telecom ranges. We will demonstrate how
the dispersion relation of propagating SPP modes can be tuned by
introducing a periodic modulation in the metal surface in a length
scale much smaller than the wavelength ($\lambda$). Even more, we
will show how this modulation could also ``build-up'' tightly
confined SPP modes in structures where these modes were not
supported in the non-corrugated case.

\begin{figure}
\centerline{\includegraphics[width=8.5cm]{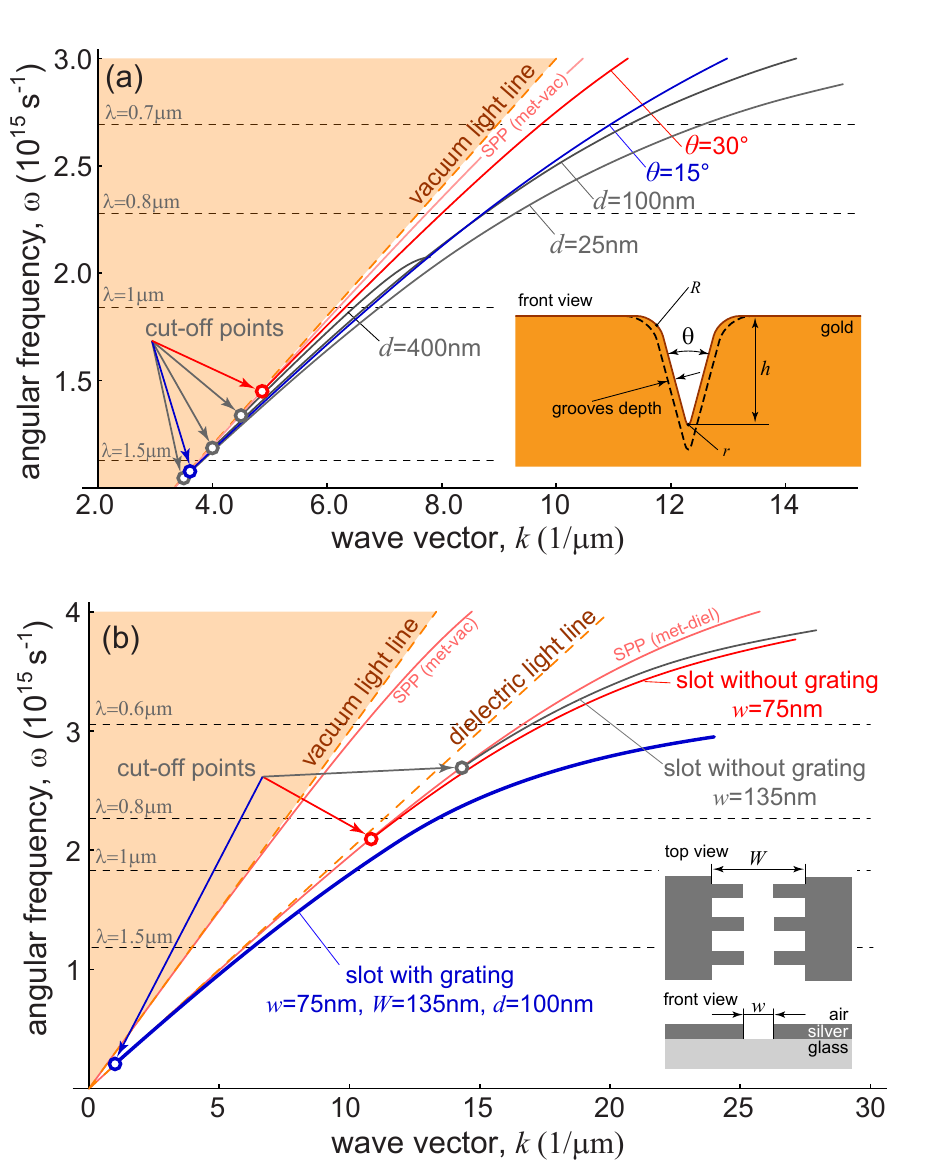}}
\caption{\label{ChannelSlot} (a) Dispersion curves for CPPs. Red
and blue curves represent those for a CPP mode of the V-groove
without corrugation for two different angles, $\theta=30^0$ and
$\theta=15^0$. The depth of the V-groove is $h=1.1 \,\mu$m and the
radii of curvature at the edges are $r=10\,$nm and $R=100\,$nm.
Grey lines display the dispersion curves for CPP modes of the
corrugated V-groove ($\theta=30^0$) where the depth of the grating
is fixed at $30\,$nm and three different periods are studied:
$d=400\,$nm, $d=100\,$nm and $d=25\,$nm. Inset shows the geometry.
(b) Dispersion curves for slot waveguide modes. Red and grey
curves are those for the slot without grating. Blue curve is the
dispersion curve for a corrugated slot. The period of the grating
is $d=100\,$nm.}
\end{figure}

First we consider a SPP mode that is bound to and propagates along
a finite V-shaped groove milled in a metal film, the so-called
channel plasmon polariton (CPP). The propagation characteristics of
these CPP modes have been studied both theoretically~\cite{Maradudin2002,Pile2004,Moreno2006} and experimentally~\cite{Bozhevolnyi2005} during the last years, and even prototype
CPP-based photonic circuits have been already fabricated~\cite{Bozhevolnyi2006}. In Fig.~\ref{ChannelSlot}(a) we plot the
dispersion relation (frequency versus parallel momentum) of the
fundamental CPP mode for a V-groove whose geometrical parameters
are taken from experiments~\cite{Bozhevolnyi2007}. The metal
considered is gold and its frequency-dependent dielectric function
is taken from~\cite{Vial2005}. All the numerical calculations
presented in this paper have been performed using a Finite Element
Method (FEM). Due to the finite depth of the grooves, this CPP
mode presents a cutoff wavelength that, for the chosen set of
geometrical parameters (see caption of Fig.~\ref{ChannelSlot}), appears at around
$1.3\,\mu$m. Grey curves in Fig.~\ref{ChannelSlot}(a) illustrate
the modification of the CPP dispersion relation induced by a
sub-$\lambda$ periodic modulation in the V-groove. The main effect
of the corrugation is to shift the cutoff wavelength of the CPP
mode to longer wavelengths, reaching a value of $1.6\,\mu$m for
the shortest period analyzed, $d=25\,$nm. Also, as its dispersion
relation departs more from the light line, the CPP for the
corrugated V-groove becomes more localized than the {\it pure}
one.

Our finding that cutoff wavelength shifts to longer wavelengths in corrugated V-grooves could explain why in the experiments there exists a propagating CPP mode in the wavelength range between 1424-1640~nm~\cite{Bozhevolnyi2005}, despite the fact that calculations for a non-corrugated V-groove with the same geometrical parameters ($\theta=25$ degrees) predict a cutoff wavelength of 1440~nm~\cite{Moreno2006}. Scanning electron microscope images reveal the presence of a weak periodic modulation in V-grooves fabricated with focused ion beam techniques.  When a shallow corrugation with a period $d=25$~nm is now introduced in the V-groove, our calculations show that the cutoff wavelength moves from 1440~nm to 1750~nm, larger than the wavelength range analyzed in the experiments.

A much stronger effect associated with a longitudinal
sub-$\lambda$ periodic corrugation is seen for another type of SPP
modes: a slot SPP mode that propagates along a gap carved between
two metal plates. Here we consider a waveguide structure that is
composed of a slot perforated on a thin silver film deposited on
top of a glass substrate ($\varepsilon=2.25$), see inset of Fig.~\ref{ChannelSlot}(b). The main geometrical parameter that controls the dispersion
relation of these modes is the width of the slot, $w$. Figure~\ref{ChannelSlot}(b) shows two different non-corrugated
structures, $w_1=75\,$nm (red line) and $w_2=135\,$nm (grey line).
The geometry of the corrugation is chosen so the distance between
two deepest points, $W$, is equal to $w_2$ whereas the minimal
distance is equal to $w_1$, see inset of
Fig.~\ref{ChannelSlot}(b). The introduction of a periodic
modulation has a dramatic effect on the dispersion relation. One
could naively expect that the resulting dispersion curve would be
located between those for the non-corrugated cases. However, the
band for the corrugated slot departs strongly from those two
curves, resulting in a much longer cutoff wavelength and larger
confinement. Then, our results clearly show that a sub-$\lambda$
periodic modulation could also be used to improve the guiding
properties of slot waveguide modes.

\begin{figure*}
\centerline{\includegraphics[width=17.5cm]{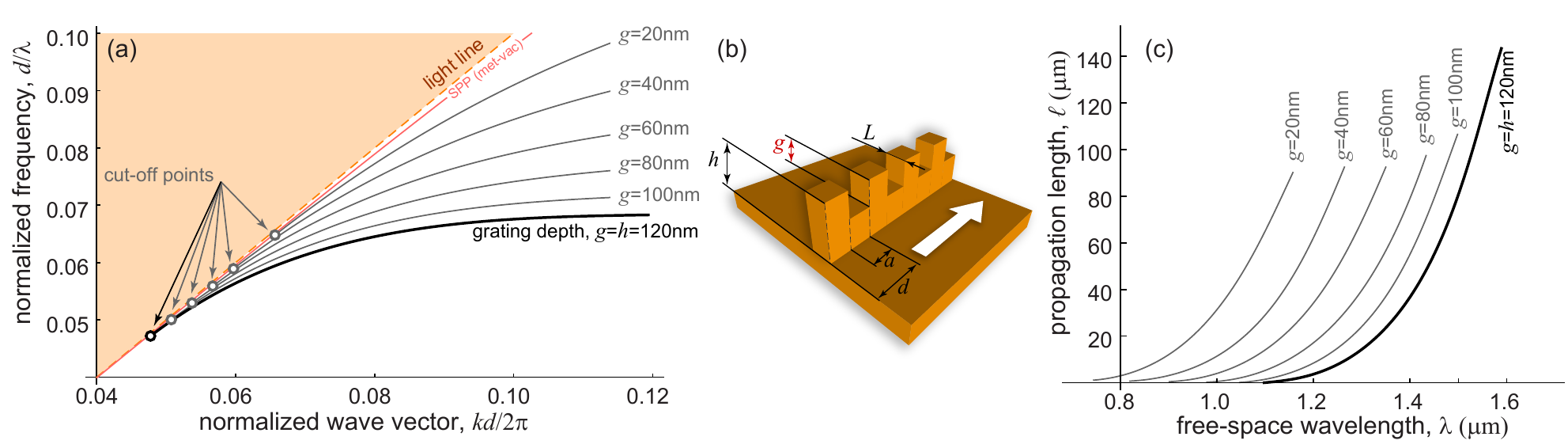}}
\caption{\label{Grating} Creation of a geometrically-induced ridge
SPP mode. (a) Dispersion curves for SPP modes running on a
corrugated ridge with different modulation depths. (b) Geometry of
the ridge structure with a modulation: corrugation period
$d=75\,$nm, width of the ridge $L=37.5\,$nm, height of the ridge
$h=120\,$nm and the grooves width $a=d/2=37.5\,$nm. (c)
Corresponding propagation lengths for the cases analyzed in panel
(a).}
\end{figure*}

Now we demonstrate that a sub-$\lambda$ periodic modulation could
indeed create a SPP mode in structures where, without corrugation,
laterally confined SPP modes are not supported. As an example we
consider a metallic ridge (height $h$ and width $L$) placed on top
of a substrate made of the same metal, see Fig.~\ref{Grating}(b). This
structure does not support the propagation of SPP modes
transversally localized. In Fig.~\ref{Grating}(a) we render the
dispersion relations of the geometrically-induced SPP modes that
emerge when a sub-$\lambda$ periodic modulation is introduced into
a gold ridge. The six curves correspond to different values of the
modulation depth, $g$. As clearly seen in the figure, even the
weakest modulation ($g=20\,$nm) is able to create a laterally
confined SPP mode (i.e., the dispersion curve is lower than the
SPP-curve for the flat metal surface). When the depth is enlarged,
the dispersion relation further departs from the light line,
increasing the mode localization. Accompanying this movement, the
cutoff frequency shifts to lower frequencies. The increase in the
mode localization also affects the propagation length of these
geometrically-induced SPP modes. For a fixed wavelength, the
transversally confined mode decays faster for a larger $g$, see
Fig.~\ref{Grating}(c).

\begin{figure}
\centerline{\includegraphics[width=8.5cm]{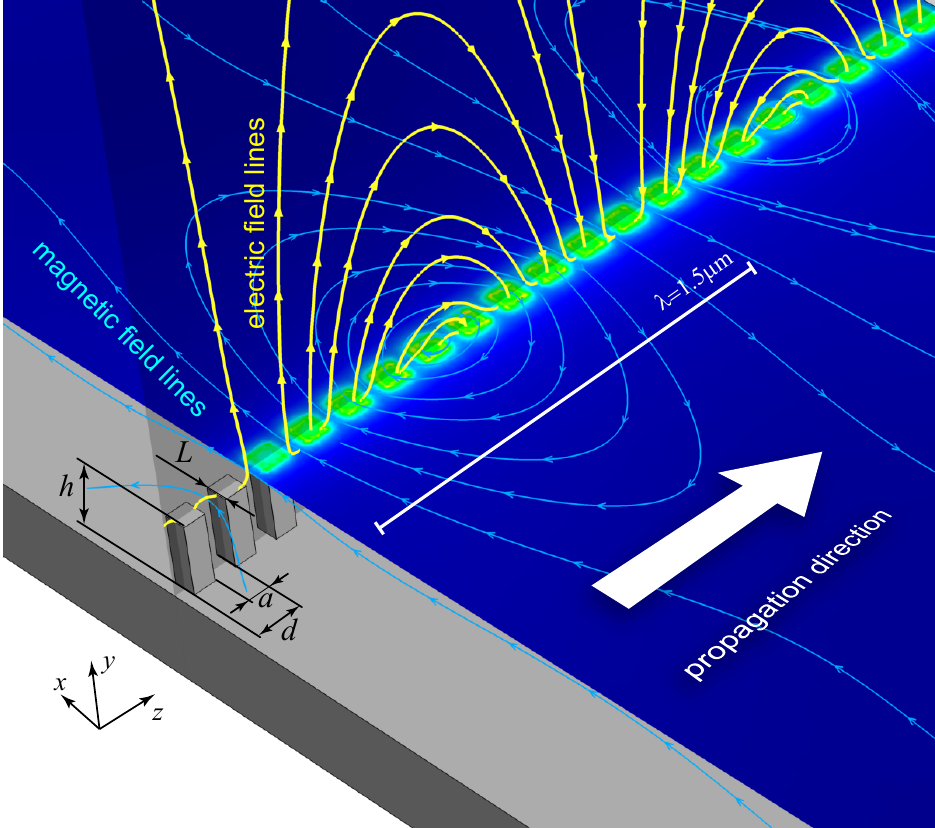}}
\caption{\label{DominoField} Electric and magnetic fields
associated with a domino plasmon polariton mode. Yellow (blue)
lines represent the electric (magnetic) vector field. Electric
field intensity map is evaluated on a horizontal plane located on
top of the domino structure. The geometrical parameters of the
domino structure are: modulation period $d=\,75$nm, lateral width
of the ridge $L=37.5\,$nm, height $h=120\,$nm and the grooves'
width $a=d/2$. The operating wavelength is $\lambda=1.5\mu$m. The
definition of the cartesian axes is also depicted.}
\end{figure}

In the case where the grating depth is equal to the height of the
slab, the geometry resembles a 1D chain of metallic box-shaped
particles placed on top of a metal film. From now on, we name the
mode supported by this structure as \textit{Domino
Plasmon-Polariton} (DPP). This DPP mode has a high electric field
localization near the top part of the domino structure, see
Fig.~\ref{DominoField}. In this figure the horizontal slice
renders the electric field intensity evaluated in a $xz$-plane
that is parallel to the metal substrate and located~$5\,$nm above
the domino's top face. The intensity also presents a strong
subwavelength confinement in the transversal direction. Regarding
the vectorial nature of the EM-fields associated with a DPP, the
electric field has mainly $y$ and $z$ components (see yellow lines
in the vertical plane of Fig.~\ref{DominoField}) whereas the magnetic field has
predominant $x$ and $z$ components (see blue lines in the
horizontal plane of Fig.~\ref{DominoField}).

It is worth discussing the differences between the DPP modes
described above and the plasmon modes supported by 1D arrays of
metal nanoparticles placed on top of a dielectric film~\cite{Krenn1999,Maier2003}. In this last case, the near-field
coupling between the localized plasmon modes of the metal
nanoparticles leads to the formation of a very flat plasmon band
characterized by a deep sub-$\lambda$ confinement but short
propagation length. However, in the case of DPP modes, the
presence of the metal substrate results in the emergence of a
``polaritonic'' part in the dispersion relation that runs close to
the SPP band of the flat surface [see Fig.~\ref{Grating}(a)]. Accordingly, DPP modes operating within this polaritonic regime posses a very long propagation length but, as expected, are much less confined than the modes supported by a chain of metal nanoparticles.

\begin{figure*}[!htb]
\centerline{\includegraphics[width=17.5cm]{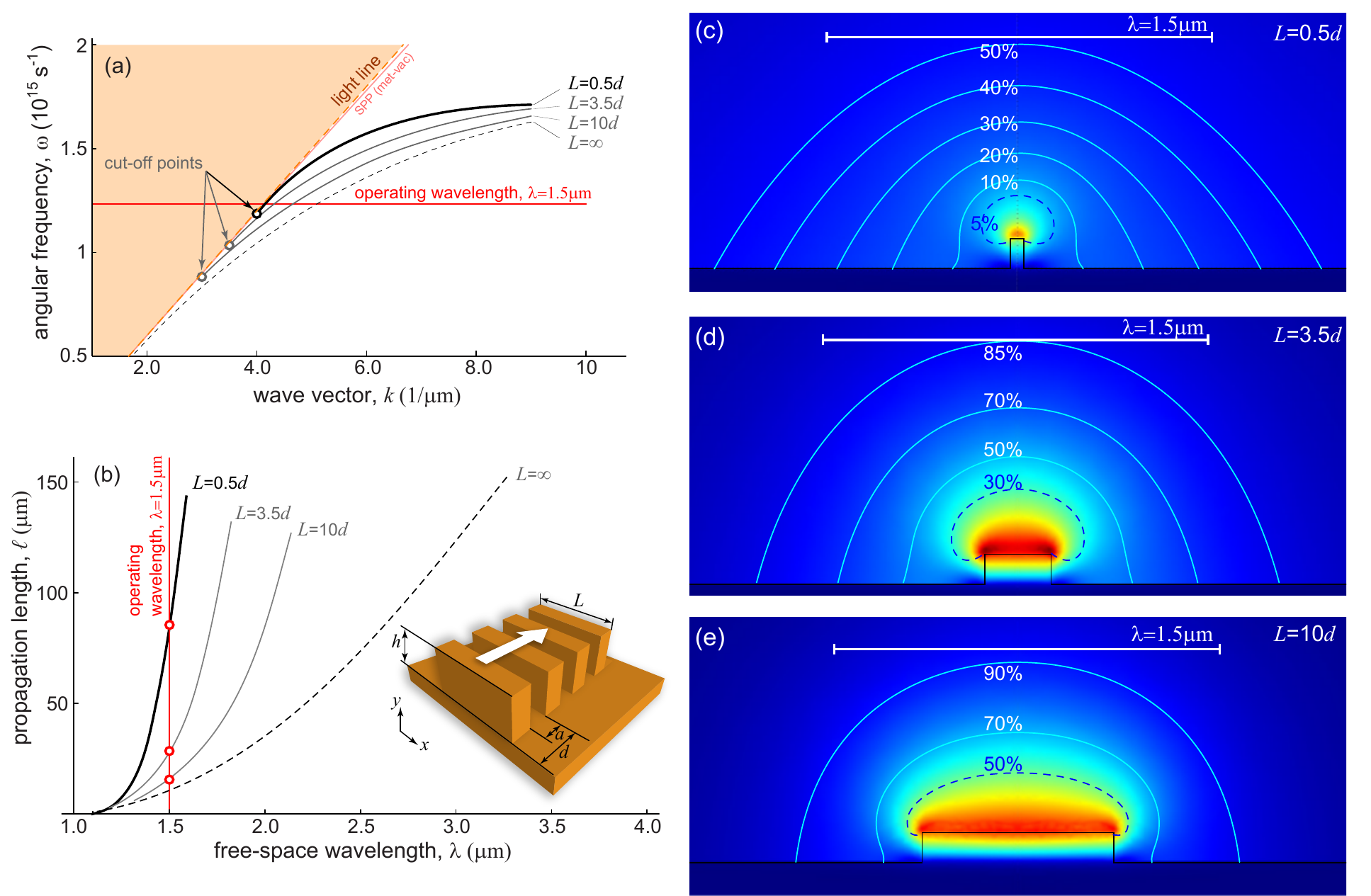}}
\caption{\label{DominoWidth} Dependence of the DPP propagation
characteristics with the domino width. (a) Dispersion curves of
the DPP modes for four different widths $L$ of the structure. The
geometrical parameters of the domino structure are the same as in
previous figures. (b) Propagation length versus free-space
wavelength for the four structures analyzed in panel (a). (c-e)
Electric field intensity distributions evaluated at $\lambda=1.5
\,\mu$m in a plane perpendicular to the propagation direction for
three different values of $L$. The blue lines show several energy
isocurves: the percentage measures the amount of energy (with
respect to total one) carried through the area bounded by the
corresponding isocurve.}
\end{figure*}

The propagation properties of a DPP mode depend on the lateral
size ($L$) of the structure. By varying this dimension we could
find optimal properties of the DPP mode, like enhanced field
confinement and/or longer propagation length. Figure~\ref{DominoWidth}(a) shows the DPP dispersion curves for four
different values of $L$. As this width is reduced from $L=\infty$
to $L=0.5d=37.5\,$nm, the cutoff frequency increases and, for a
fixed frequency, the wavevector is smaller for narrower dominos.
Accordingly, the DPP propagation length for very narrow dominos is
larger than that for a wider one (see Fig.~\ref{DominoWidth}(b)).
Interestingly, even for a domino width of only $L=d/2=37.5\,$nm
the propagation length at $\lambda=1.5\mu$m is around $80\,\mu$m,
a value that is larger than those predicted for other types of SPP
modes that present sub-$\lambda$ confinement like CPPs~\cite{Moreno2006} or wedge plasmon polaritons~\cite{Moreno2008}.
When the metal behaves as a perfect electrical conductor, our
calculations show that the dispersion relation of the DPP mode is
almost independent of $L$, paving the way to several interesting
applications for the use of DPPs in the terahertz regime~\cite{Martin-Cano2009}.

Figures~\ref{DominoWidth}(c-e) render the electric field intensity
distribution for three DPPs in a plane perpendicular to the
propagation direction located at the center of the gap between two
dominos. The lateral sizes analyzed are $L=0.5d$, $L=3.5d$ and
$L=10d$ and the operating wavelength is $\lambda=1.5\,\mu$m. The
color scale is the same for the three distributions. As expected
from the dispersion curves, the DPP mode is more confined for the
wider domino, $L=10d$. When the domino width is decreased, the EM
energy is carried within a bigger volume. This is illustrated in
Figs.~\ref{DominoWidth}(c-e) by representing the energy isocurves
for the three different dominos. The percentage associated with
each isocurve measures the amount of EM energy (with respect to
the total one) carried out through the area inscribed into the
curve. This is the usual way to define the modal size of a
waveguide mode~\cite{Oulton2008}. However, a closer look at the
intensity distributions shows that the electric field in the case
$L=3.5d$ presents a stronger field localization than the $L=10d$
or $L=0.5d$ cases. Nevertheless, a comparison of electric field
distributions with the operating wavelength (white bars) in Fig. 4
illustrates the strong sub-$\lambda$ transversal confinement
associated with the propagation of DPP modes.

In conclusion, we have shown that the application of the spoof
plasmon concept in the optical and telecom regimes allows
tailoring of the guiding properties of SPP modes. Moreover,
we have also demonstrated that the same approach leads to the
emergence of guided modes in geometries where conventional SPPs
are not supported. We have illustrated this finding by analyzing
the so-called Domino Plasmon Polaritons, which exhibit an
excellent trade-off between lateral confinement and propagation
length.

This work was sponsored by the the Spanish Ministry of Science
under projects MAT2009-06609-C02 and No. CSD2007-046-NanoLight.es.

\end{document}